\documentclass[
preprint,
 amsmath,amssymb,
 aps,
prb,
floatfix,
]{revtex4-1}

\usepackage{graphicx}
\usepackage{dcolumn}
\usepackage{bm}
\usepackage{epsfig}
\usepackage{color}
\usepackage{amsmath}
\usepackage{hyperref}
\usepackage[english]{babel}

\hypersetup{colorlinks=true, urlcolor=blue, citecolor=blue, linkcolor=blue, anchorcolor=blue}  

\newcommand{\be}{\begin{equation}}
\newcommand{\ee}{\end{equation}}
\newcommand{\ba}{\begin{align}}
\newcommand{\ea}{\end{align}}

\begin{document}
	
\title{Higher-order topological insulators in synthetic dimensions}
	
\author{Avik Dutt}
\author{Momchil Minkov}
\author{Shanhui Fan}
\affiliation{Ginzton Laboratory and Department of Electrical Engineering, Stanford University, Stanford, CA 94305, USA}

\begin{abstract}
Conventional topological insulators support boundary states that have one dimension lower than the bulk system that hosts them, and these states are topologically protected due to quantized bulk dipole moments. Recently, higher-order topological insulators have been proposed as a way of realizing topological states that are two or more dimensions lower than the bulk, due to the quantization of bulk quadrupole or octupole moments. However, all these proposals as well as experimental realizations have been restricted to real-space dimensions. Here we construct photonic higher-order topological insulators (PHOTI) in synthetic dimensions. We show the emergence of a quadrupole PHOTI supporting topologically protected corner modes in an array of modulated photonic molecules with a synthetic frequency dimension, where each photonic molecule comprises two coupled rings. By changing the phase difference of the modulation between adjacently coupled photonic molecules, we predict a dynamical topological phase transition in the PHOTI. Furthermore, we show that the concept of synthetic dimensions can be exploited to realize even higher-order multipole moments such as a 4$^{th}$ order hexadecapole (16-pole) insulator, supporting 0D corner modes in a 4D hypercubic synthetic lattice that cannot be realized in real-space lattices.
\end{abstract}

\maketitle

\section*{Introduction}
A conventional topological insulator in 2D and 3D supports gapless edge states and surface states respectively that are protected against local perturbations by the nontrivial topology of the bulk. The existence of these gapless states, which have one dimension lower than the bulk that hosts them, is guaranteed by the bulk-boundary correspondence. Recently, the concept of higher-order topological insulators (HOTIs) has been proposed to generalize this bulk-boundary correspondence, revealing the existence of topological states that are two or more dimensions lower than the bulk. In general, an $n$-th order topological insulator in $D$-dimensions supports $(D-n)$-dimensional topological boundary modes of codimension $n$. The first such prediction was of zero-dimensional zero-energy corner states in a second-order topological insulator whose edge states are gapped, and the existence of these zero-energy corner states is guaranteed by a quantized bulk quadrupole moment~\cite{benalcazar_quantized_2017}. Following closely on the heels of this theoretical prediction, quadrupole HOTIs have been experimentally realized in several systems, including bismuth~\cite{schindler_higher-order_2018-1}, mechanical metamaterials~\cite{serra-garcia_observation_2018}, acoustics~\cite{xue_acoustic_2019, ni_observation_2019}, electrical circuits~\cite{imhof_topolectrical-circuit_2018}, and photonics~\cite{mittal_photonic_2019}. However, both the theoretically proposed and experimentally demonstrated HOTIs have been restricted to real-space dimensions, that is, spatially periodic lattices.

\begin{figure}
\includegraphics[width=.96\textwidth]{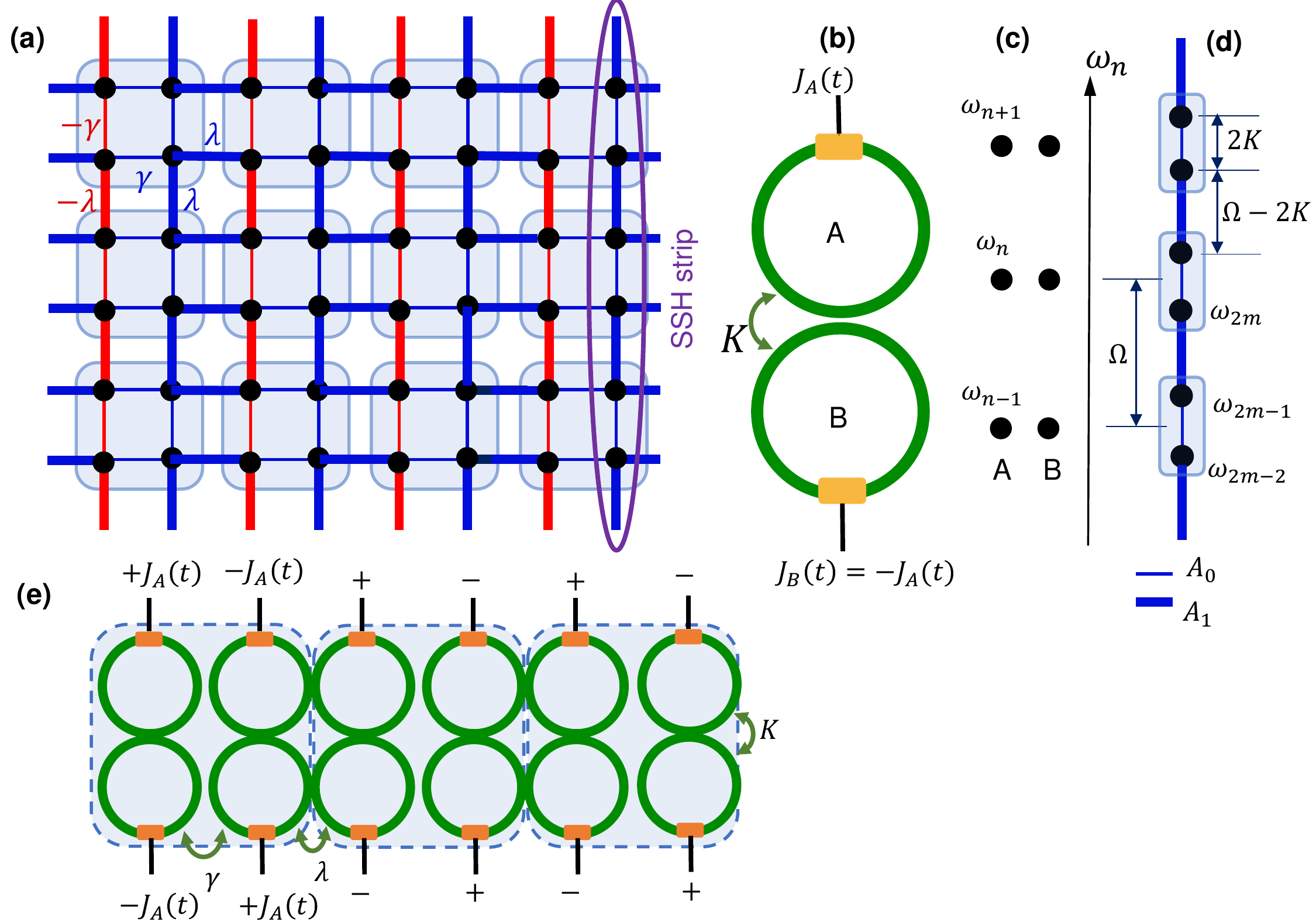}
\caption{\linespread{1.2}\selectfont{}(a) Tight-binding lattice of a quadrupole topological insulator. Blue and red lines represent positive and negative real values of the coupling strength. Thin and thick lines represent intra-cell and inter-cell coupling strengths of magnitude $\gamma$ and $\lambda$ respectively. Each vertical column is an SSH strip with the same sign of coupling strength throughout, and adjacent vertical lines have a phase offset of $\pi$ in the coupling strength. (b) Implementation of the SSH strip in the synthetic frequency dimension using modulated coupled rings (photonic molecule), with $J_A(t)$ as in Eq.~\eqref{eq:mod}. (c) Mode structure of the system in (b) with sets of frequency modes separated by the FSR $\Omega$, in the absence of modulation and coupling. (d) Mode spectrum of the coupled ring system or photonic molecule. Each set is split by a frequency difference equal to the coupling constant between the rings, $2K$. The modulation introduces coupling between the supermodes. (e) Several of the synthetic SSH strips in (b) can be evanescently coupled with alternate coupling strengths $\gamma$ and $\lambda$ to realize the quadrupole HOTI lattice of (a). We note that the modulation pattern of the unit cell of four rings has a quadrupole nature.}
\label{fig:1_lattice}
\end{figure}

In contrast to real-space dimensions, synthetic dimensions are formed by coupling internal degrees of freedom. These can be, for example, the frequency, arrival time, or orbital angular momentum of photons, or the spin of ultracold atoms~\cite{yuan_synthetic_2018-1, ozawa_topological_2019}. Introducing coupling between these degrees of freedom then allows the study of higher-dimensional physics on lower dimensional structures~\cite{jukic_four-dimensional_2013, boada_quantum_2012, lustig_photonic_2019}. 
A prime focus of research in synthetic dimensions has been the pursuit of conventional topological phases in simple structures, such as the study of the 2D quantum Hall effect in a 1D real-space array~\cite{yuan_photonic_2016, ozawa_synthetic_2016, luo_quantum_2015, mancini_observation_2015, stuhl_visualizing_2015}, or the study of 3D topological physics in a 2D planar array~\cite{lin_photonic_2016, lin_three-dimensional_2018}. Additionally, researchers have studied two or more simultaneous synthetic dimensions to implement higher-dimensional physics in essentially 0D systems~\cite{yuan_synthetic_2018, martin_topological_2017, dutt_single_2019, yuan_photonic_2019, reimer_high-dimensional_2019}. Since the concept of synthetic dimensions is well suited to study topological physics in high-dimensional lattices, a natural question is whether higher-order topological insulators can be realized in synthetic space.

Here we answer this question in the affirmative by constructing a photonic higher-order topological insulator (PHOTI) in synthetic dimensions. Our system consists of pairs of ring resonators that are mutually coupled to form an array, realizing a 1D chain of ``photonic molecules"~\cite{spreeuw_classical_1990, zhang_electronically_2019}. By antisymmetrically modulating the two rings in a photonic molecule at the frequency spacing between the modes of the ring, we realize a lattice along the synthetic frequency dimension. A 1D array of modulated photonic molecules realizes a quadrupole PHOTI in the synthetic frequency dimension, in which we show the excitation of topologically nontrivial corner modes. By switching the phase difference of the modulation between adjacent photonic molecules, we show a phase transition from the topologically protected phase with a nonzero quantized quadrupole moment to a phase with zero quadrupole moment. Additionally, we propose, for the first time, a hexadecapole (16-pole) insulator with topologically nontrivial corner modes by leveraging synthetic dimensions to create a 4D hypercubic lattice that cannot be realized in real-space lattices. Our work illustrates the potential of using the concept of synthetic dimensions for exploring exotic new phases including very-high-order topological insulators in high dimensions.

\section*{Results}

\subsection*{Quadrupole PHOTI}
Consider the lattice in Fig.~\ref{fig:1_lattice}(a) for a quadrupole higher-order topological insulator (HOTI)~\cite{benalcazar_quantized_2017}. Each vertical 1D strip in this lattice is a Su-Schrieffer-Heeger (SSH) strip~\cite{su_solitons_1979}, with alternating values of the coupling $\gamma$ and $\lambda$ representing the intra-cell and inter-cell hopping strengths respectively. The signs of the coupling along each vertical strip are the same, and adjacent lines have a $\pi$ phase difference between the coupling. 
All the horizontal couplings have real positive values. Thus, there is a $\pi$ magnetic flux through each of the plaquettes.  

We show that the model of Fig.~\ref{fig:1_lattice}(a) can be realized with the use of the concept of a synthetic dimension. To construct each SSH strip of the quadrupole HOTI, we use a pair of mutually coupled identical ring resonators A and B, each with an electro-optic modulator, as shown in Fig.~\ref{fig:1_lattice}(b). Such a pair of photonic cavities, with or without modulation, is often called a ``photonic molecule"~\cite{spreeuw_classical_1990, zhang_electronically_2019, majumdar_cavity_2012, galbiati_polariton_2012, dutt_tunable_2016, nakagawa_photonic_2005} in analogy with a diatomic molecule. Each individual ring supports longitudinal cavity resonances at $\omega_n = \omega_0 + n\Omega$, separated by the free spectral range (FSR) $\Omega/2\pi = v_g/L$, where $v_g$ is the group velocity of light in the ring, and $L$ is its length [Fig.~\ref{fig:1_lattice}(c)]. In forming the photonic molecule, the two modes of the two individual rings at the same frequency $\omega_n$ hybridize into symmetric and antisymmetric supermodes, with frequencies $\omega_{n-} = \omega_n - K \equiv \omega_{2m}$ and $\omega_{n+} = \omega_n + K \equiv \omega_{2m+1}$ respectively, where $K$ is the coupling strength between the rings [Fig.~\ref{fig:1_lattice}(d)]. We neglect dispersion in the ring-to-ring coupling strengths. The frequencies of the photonic molecule in the basis of symmetric and antisymmetric supermodes thus form a strip with alternating spacings $2K$ and $\Omega-2K$. By choosing a modulation of the form,
\be
J_A(t, \phi) = A_0 \cos (2Kt+\phi) + A_1 \cos[(\Omega-2K)t + \phi];\  J_B(t, \phi) = -J_A(t, \phi) \label{eq:mod}
\ee
one can form a synthetic frequency dimension with alternating coupling strengths $A_0$ and $A_1$ (see Supplementary Materials). The antisymmetric modulation $J_B = -J_A$ is necessitated by the opposite symmetry of the two supermodes at adjacent frequencies. Throughout the paper, we assume $A_0, A_1 \ll K < \Omega/2$ so that the rotating wave approximation is valid.

Next, to form the full 2D lattice for the quadrupole insulator in Fig.~\ref{fig:1_lattice}(a), we form a lattice of the pairs of cavities as described above, with alternate coupling strengths $\gamma$ and $\lambda$ determined by the respective coupling gaps between the nearest neighbor cavities along the horizontal axis [Fig.~\ref{fig:1_lattice}(e)]. This system is described by a two-dimensional synthetic space with a real space axis and a synthetic frequency axis. The signs of the modulation between adjacent cavity pairs are switched ($\phi=0$ and $\phi=\pi$ in Eq.~\eqref{eq:mod}) to implement a flux of $\pi$ per plaquette in the two-dimensional lattice~\cite{fang_realizing_2012, yuan_photonic_2016}.
Interestingly, the signs of the modulation in each unit cell of this array of paired resonators follow a quadrupole pattern, as seen in Fig.~\ref{fig:1_lattice}(e). The strength of the modulation is chosen to satisfy $A_0=2\gamma$, $A_1 = 2\lambda$. 

\begin{figure}
\includegraphics[width=.96\textwidth]{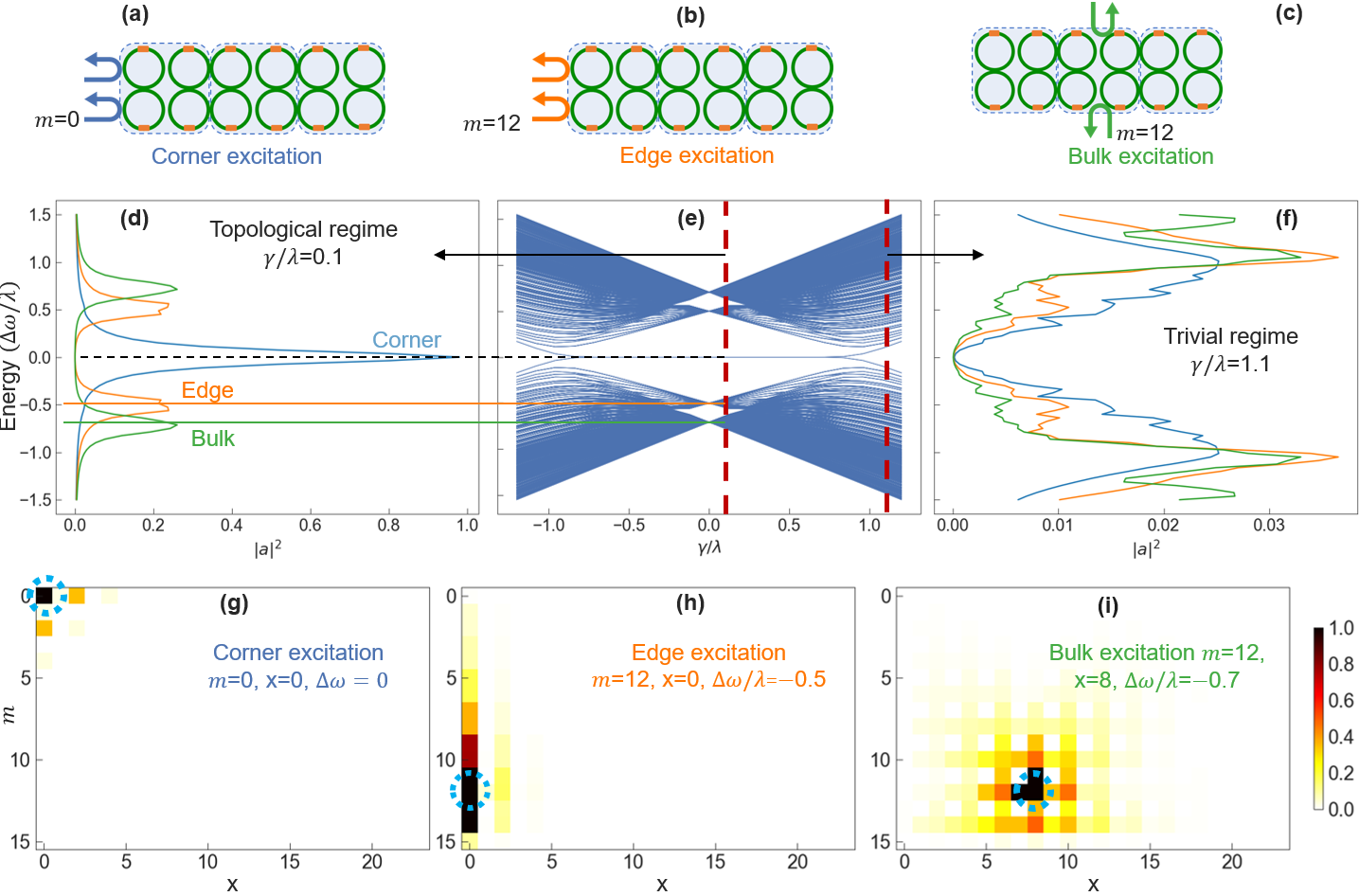}	
\caption{\linespread{1.2}\selectfont{}
Energy eigenspectrum, edge modes and corner modes of the synthetic dimension quadrupole HOTI. {\bf (a)-(c)} Schematics for selective excitation of corner, edge and bulk modes in the photonic molecule array using external waveguides.  
{\bf (d)} Intensity $|a|^2$ in the excited rings at various input frequency detunings $\Delta \omega$, for excitations as indicated in (a)-(c). Peaks appear  at input frequency detunings that correspond to the positions of the corner, edge and bulk modes in (e). {\bf (e)} Energy eigenspectrum for a finite lattice of the quadrupole HOTI. For $|\gamma/\lambda|<1$, the system exhibits topologically protected corner modes pinned to zero energy. For $|\gamma/\lambda|>1$, the corner modes cease to exist. {\bf (f)} Same as (d), but in the topologically trivial regime $\gamma/\lambda = 1.1$. No peaks are seen in the bandgap as the corner and edge modes cease to exist. Bottom row: cavity field intensity for exciting the finite lattice at the corner {\bf (g)}, edge {\bf (h)} and bulk {\bf (i)} for $\gamma/\lambda=0.1$. No such corner or edge localized modes were seen in the trivial phase on the same excitations, for $|\gamma/\lambda| >1$. Color scale in the bottom row indicates steady state distribution of the modes in the lattice. Blue dashed circles denote the lattice site that is excited in each case.}
\label{fig:2_excitation}
\end{figure}

\subsection*{Excitation of corner modes}
The hallmark of the quadrupole HOTI model as described in Fig.~\ref{fig:1_lattice}(a) is the existence of fourfold degenerate zero-energy corner modes with codimension 2, while the edge modes are gapped, for $|\gamma/\lambda|<1$. In our implementation with an array of modulated photonic molecules, as shown in Fig.~\ref{fig:1_lattice}(e), these midgap corner modes can be excited by coupling external waveguides to the array [Fig.~\ref{fig:2_excitation}(a)-(c)]. The demonstration of these corner modes then serves as an indication that we have indeed constructed a quadrupole photonic HOTI in synthetic space. Since these corner modes only exist in a finite lattice, we choose $M_\omega = 16$ sites (8 pairs of supermodes) along the frequency axis for our calculations. Such a termination of the frequency axis can be achieved by strongly coupling a ring with a radius $M_\omega$ times smaller than the main rings, to induce a strong local change in the FSR every $M_\omega$ modes~\cite{yuan_photonic_2019}. Alternatively, one can engineer the dispersion of the ring waveguide to strongly perturb the FSR beyond the $M_\omega$ modes~\cite{yuan_photonic_2016}, which makes the modulation in Eq.~\eqref{eq:mod} off-resonant beyond the finite lattice formed by these $M_\omega$ modes.

Fig.~\ref{fig:2_excitation}(d) shows the results of exciting the photonic molecule array in the topological phase $\gamma/\lambda=0.1$. We can observe corner, edge and bulk modes by exciting suitable rings at the appropriate frequency ($\omega_{\rm in} = \omega_{m} + \Delta\omega$), where $m$ denotes the desired frequency mode and $\Delta\omega$ maps to the quasi-energy when this time-modulated system is treated as a Floquet system. Note that selective excitation of a single site in the synthetic lattice requires the excitation of two rings either in phase (+ mode, $m$ even) or out of phase ($-$ mode, $m$ odd), as the modes in each photonic molecule are symmetric and antisymmetric combinations of the isolated ring modes [Fig.~\ref{fig:2_excitation}(a)-(c)]. For corner mode excitation, we choose the leftmost pair of rings with an excitation frequency $\omega_{\rm in} = \omega_{m=0} + \Delta \omega$.  We observe a peak for $\Delta\omega = 0$, because the midgap corner modes are pinned to zero energy, which is consistent with the eigenspectrum shown in Fig.~\ref{fig:2_excitation}(e). The corresponding field distribution in the synthetic lattice for $\Delta\omega = 0$ is shown in Fig.~\ref{fig:2_excitation}(g), which exhibits strong localization at the corner. For edge mode excitation, we choose the same pair of rings but change $\omega_{\rm in}$ to $\omega_{m=12} + \Delta\omega$, and observe peaks for $\Delta\omega/\lambda \approx \pm 0.5$. Between the peaks there is a lack of output amplitude, which indicates that the edge modes are gapped. For bulk excitation, we choose a pair of rings in the center of the array, and observe peaks at $\Delta\omega/\lambda \approx \pm 0.70$, in accordance with the eigenspectrum in Fig.~\ref{fig:2_excitation}(e). The corresponding field distributions in the synthetic lattice for exciting the edge and bulk modes at their respective detunings $\Delta\omega$ are shown in Fig.~\ref{fig:2_excitation}(h) and (i). By contrast, in the trivial regime $\gamma/\lambda=1.1$, we see no midgap peaks because the corner modes cease to exist -- in fact, there are no modes in the bandgap even for a finite lattice [Fig.~\ref{fig:2_excitation}(f)].

\begin{figure}[h]
\includegraphics[width=1.05\textwidth]{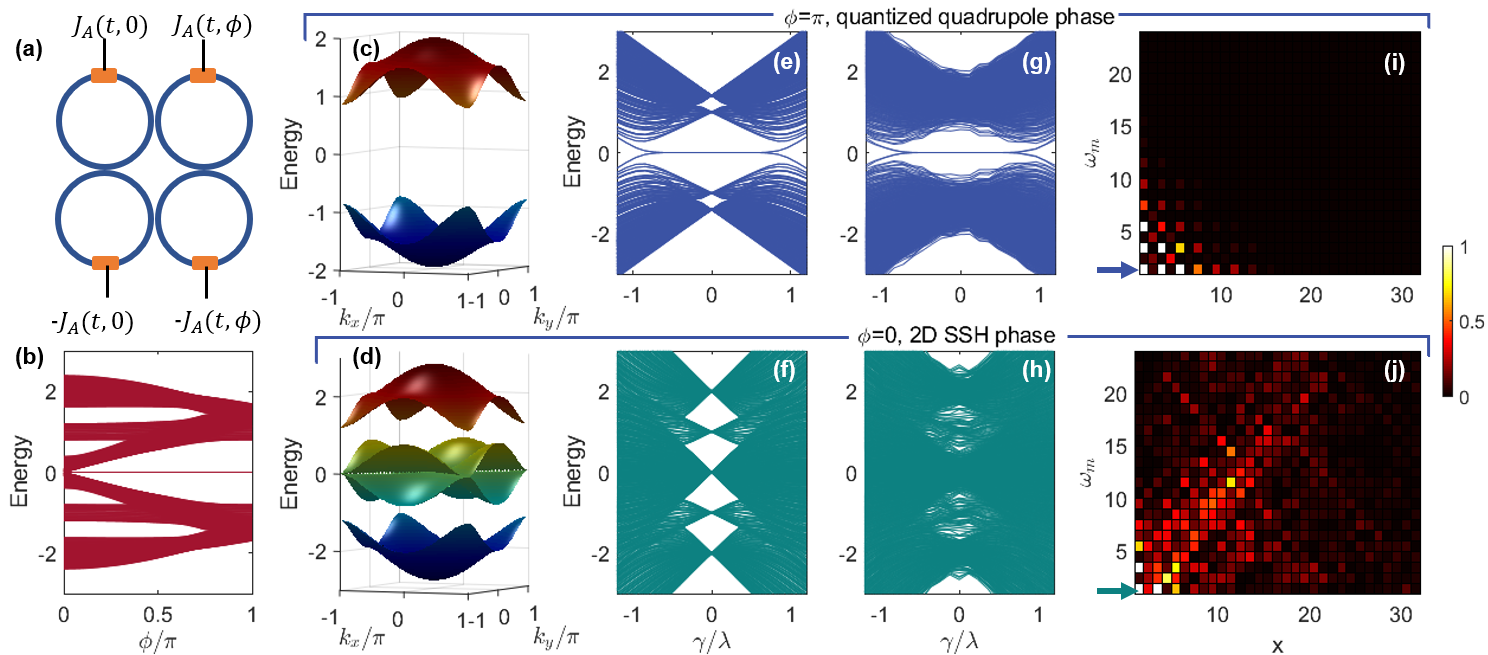}
\caption{\linespread{1.2}\selectfont{}
Topological phase transition in the synthetic PHOTI lattice by tuning the phase of modulation. (a) Unit cell consisting of two photonic molecules with a phase difference $\phi$ between the modulations [Eq.~\eqref{eq:mod}]. (b) Energy eigenspectrum for a finite lattice with open boundaries in both real dimension $x$ and synthetic frequency dimension $m$ for various phases $\phi$, with $\gamma/\lambda = 0.20$. For $\phi=\pi$, there is a bandgap that hosts topologically protected corner modes pinned to zero energy. For $\phi=0$, no such bandgap exists at zero energy, as confirmed by the bulk band structure in (d). (c) and (d) show the bulk band structures for quadrupole HOTI ($\phi=\pi$) and the 2D SSH model ($\phi=0$), respectively. All energies are in units of $\lambda$. In (c), both bands are doubly degenerate. (e), (f) Energy eigenspectrum of an ideal finite lattice for $\phi=\pi$  and $0$ respectively, without disorder in the couplings. Though corner modes exist in both cases, for $\phi=0$, they spectrally overlap with the bulk bands. (g), (h) Energy eigenspectrum with normally distributed random disorder in the couplings with variance $\sigma^2=0.04$. Since the lattice with $\phi=\pi$ hosts a quantized bulk quadrupole moment, corner modes are visible in the bandgap in (g), unlike in (h). (i),(j) Steady-state field distribution in the disordered synthetic lattice sites for an excitation with zero detuning $\Delta\omega=0$ at the lowest frequency mode $m=0$ for the leftmost ring, as indicated by the arrows. For the quadrupole PHOTI ($\phi=\pi$) in the top row, the corner modes are still strongly localized in the presence of disorder. For the 2D SSH phase in the bottom row, disorder in the couplings makes the corner excitation not well localized, leaking into the bulk. Specifically, the excitation preferentially propagates at $\sim\pm$45$^\circ$ in the lattice~\cite{benalcazar_bound_2019}, because the bands in (d) touch at zero energy along the $k_x = k_y$ and $k_x = -k_y$ lines. In (c)-(j), $\gamma/\lambda=0.4$.}
\label{fig:3_phasetransition}
\end{figure}

\subsection*{Topological phase transition}
The concept of a synthetic dimension provides great flexibility in dynamically reconfiguring the hopping amplitudes and phases by changing the strengths and phases of the modulation, respectively.  We use this flexibility to show a topological phase transition between the regime with a quantized bulk quadrupole moment and a 2D SSH phase with no quadrupole moment, which occur for $\pi$ and $0$ flux per plaquette respectively. The lattice with zero flux (a 2D SSH model) possesses all the symmetries of the quadrupole insulator, namely it is invariant with respect to the translation, reflection (in x and y) and time-reversal operations. While this ensures that the bulk quadrupole moment is quantized, its value is zero. In fact, there is not even a bulk band gap at zero energy in this model, meaning that the bulk is not insulating.
Our photonic molecule array can implement such a change in flux by changing the relative phase between modulations on adjacent molecules [Fig.~\ref{fig:3_phasetransition}(a)]. In Fig.~\ref{fig:3_phasetransition}, we plot the energy eigenspectrum for various $\phi$. The bandgap remains open for intermediate values of flux $0<\phi<\pi$, but eventually closes for $\phi \rightarrow 0$. However, the quadrupole moment is not quantized for intermediate values of flux $\phi$ due to the breaking of reflection symmetry along the frequency dimension~\cite{benalcazar_electric_2017}. The bulk band structures for $\phi=\pi$ and $\phi=0$ are plotted in Fig.~\ref{fig:3_phasetransition}(c) and (d). The 2D SSH model with $\phi=0$ is not an insulator at zero energy since the bulk is not gapped for $E=0$, and although corner modes exist, they overlap spectrally with the bulk excitations. To compare the topological protection of corner modes in the quantized quadrupole phase $\phi=\pi$ and the 2D SSH phase, we introduced disorder in the couplings. The corresponding eigenspectra [Fig.~\ref{fig:3_phasetransition}(g), (h)] retain the well-separated midgap corner modes for the quadrupole phase, but not for the 2D SSH case. On exciting the corner site in the two cases, strong corner localization of the field distribution is seen for $\phi=\pi$ [Fig.~\ref{fig:3_phasetransition}(i)], but significant leakage into the bulk is seen for $\phi=0$ [Fig.~\ref{fig:3_phasetransition}(j)]. The leakage into the bulk preferentially happens along the $k_x = k_y$ and $k_x = -k_y$ directions since the central bulk subbands in (d) touch at zero energy, which is the energy where the corner modes exist for a finite lattice. 
This lack of protection of the corner modes for $\phi=0$ is expected as the system has zero bulk quadrupole moment and no band gap at zero energy. Recently, this 2D SSH model without magnetic flux has received some attention~\cite{xie_second-order_2018, benalcazar_bound_2019, ota_photonic_2019} because it can be associated with a non-zero 1D Zak phase in both directions. This is in fact what ensures the existence of the corner modes in the system without disorder. However, as we can see in Fig.~\ref{fig:3_phasetransition}(j), these corner states are not as robust as the ones of the bulk quardupole isolator phase, which is harder to achieve in real-space  due to the requirement of negative-valued couplings but is straightforward using synthetic dimensions.

An alternative way to implement a topological phase transition is to tune the ratio of intra-cell hopping $A_0$ to the inter-cell hopping $A_1$ in the synthetic frequency dimension by varying the modulation amplitude. This produces an anisotropic 2D SSH model with $\pi$ flux per plaquette, as the hoppings along the real-space axis $x$ are fixed. However, the corner modes only exist for $A_0/A_1<1$. As the modulation amplitudes are tuned from $A_0<A_1$ to $A_0>A_1$,
the one-dimensional Zak phase along the frequency dimension becomes topologically trivial, and the corner modes disappear. The Zak phase along the real dimension, however, remains non-trivial, and, upon truncation in the real space, edge modes at the boundary of real dimension still exist~\cite{xie_second-order_2018}.

\begin{figure}
\includegraphics[width=.85\textwidth]{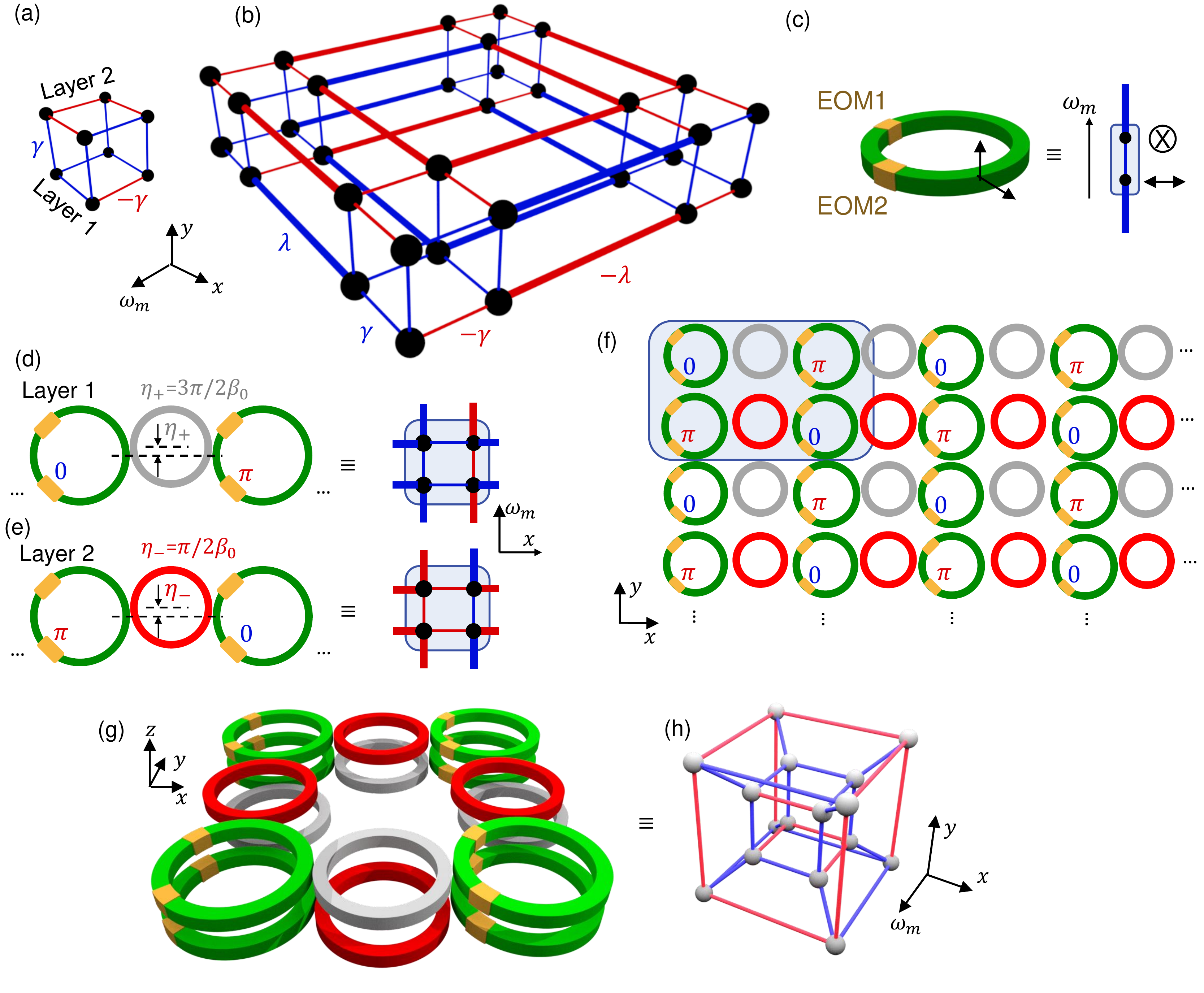} 
\caption{\linespread{1.2}\selectfont{}
Hierarchical construction of octupole and hexadecapole insulators using synthetic dimensions. {\bf (a)} Unit cell, and {\bf (b)} tight-binding model of the octupole insulator. Thin lines have coupling strength $\gamma$ and thick lines have coupling strength $\lambda$. Blue and red lines represent positive and negative coupling strengths respectively. {\bf (c)} Dipole insulator (SSH strip) formed using two polarization modes in a single resonator~\cite{spreeuw_classical_1990}. EOM1 introduces a frequency offset between the resonances of the two polarizations. EOM2 is modulated with a signal similar to Eq.~\eqref{eq:mod} to form the synthetic frequency dimension spanned by $\omega_m$. {\bf (d)} Unit cell of the quadrupole insulator formed by coupling two site rings (green) with an auxiliary link ring (grey) with a slightly smaller length, and asymmetrically placed with respect to the coupling region by an offset $\eta_+=\pi/2\beta_0$~\cite{hafezi_imaging_2013}. This implements layer 1 of the octupole unit cell in (a). Modulation phases between adjacent site rings differs by $\pi$. {\bf (e)} Implementation of layer 2 of the octupole unit cell in (a). The red link ring implements a negative coupling in real space, by having an offset $\eta_- = 3\pi/2\beta_0$. Negative-valued real-space coupling strengths are needed in our construction of octupole and hexadecapole insulators. {\bf (f)} 2D lattice of modulated rings with a synthetic frequency dimension forming the 3D octupole insulator of (b). {\bf (g)} Array of rings implementing the unit cell of the hexadecapole insulator using two layers of the octupole insulator, vertically coupled. Phases of all synthetic and real-space couplings alternate between the two layers. The vertical ring couplings are positive-valued. (h) 4D hypercubic unit cell of the hexadecapole insulator. The inner cube is realized using the bottom layer in (g), and the outer cube is realized using the top layer.}
\label{fig:4_octupole}
\end{figure}

\subsection*{Octupole and Hexadecapole insulators}
Finally, we show how the concept of synthetic dimensions can be exploited to construct PHOTIs of even higher order, such as an octupole insulator in a 3D cubic lattice and a hexadecapole (16-pole) insulator in a 4D hypercubic lattice supporting corner modes with codimension 4. 
The unit cell for the octupole insulator's cubic lattice is shown in Fig.~\ref{fig:4_octupole}(a). It consists of two layers of the unit cell for the quadrupole insulator connected by positive-valued couplings $\gamma$~\cite{benalcazar_quantized_2017}. The signs of all couplings are reversed between the two  layers [Fig.~\ref{fig:4_octupole}(a)]. The full lattice for the octupole insulator can be created by connecting such unit cells with coupling strength $\pm\lambda$, such that a $\pi$ flux is maintained in each plaquette. An example of a small finite lattice for this model is shown in Fig.~\ref{fig:4_octupole}(b). 
Regardless of the multipole order, multipole HOTIs can be viewed as composed of 1D SSH strips connected in a certain way. In Fig.~\ref{fig:4_octupole}(b) for example, a 1D strip in any direction is an SSH chain with either a positive or a negative sign for all of its couplings. The crucial characteristic for each dimension is then whether the couplings of the SSH chains flip sign when the chains are stacked in that dimension (as in the $x$- and $\omega$- dimensions of Fig.~\ref{fig:4_octupole}(b)), or if they are all of the same sign (as in the $y$-dimension). The general rule then is that the SSH chains flip sign along all but one of the dimensions.

In the construction of a quadrupole PHOTI, in connection with the experiments in Ref.~\onlinecite{dutt_single_2019}, we formed an SSH model using two rings, and utilized only one of the two polarizations that the ring can support. For constructing octupole and hexadecapole PHOTIs, since we will be using a much larger number of rings, it is of interest to reduce the numbers of rings used. Therefore, we instead construct the SSH model using only one ring resonator, but utilizing the polarization degree of freedom. For this purpose we consider a set up as shown in Fig.~\ref{fig:4_octupole}(c), where two electro-optic modulators EOM1 and EOM2 are incorporated in a single ring.  This setup was previously used in Ref.~\onlinecite{spreeuw_classical_1990} for realizing a photonic molecule but without a synthetic frequency dimension. Here, the two modes forming the SSH unit cell are the polarizations in-plane with the ring and perpendicular to the ring [Fig.~\ref{fig:4_octupole}(c)]. The splitting between the resonance frequencies of these two polarizations is proportional to the voltage applied on EOM1, and can be tuned to be $2K$, similar to Fig.~\ref{fig:1_lattice}(d). Next, using EOM2, these polarizations are coupled to each other and to the modes separated by an FSR using the modulation in Eq.~\eqref{eq:mod} with frequency components at $2K$ and $\Omega-2K$. To facilitate this coupling, the principle axes of EOM2 are at an angle of 45$^\circ$ with respect to the principle axes of EOM1.

After realizing the SSH model in a single ring, we implement the unit cell of the quadrupole insulator using two such rings, as shown in Fig.~\ref{fig:4_octupole}(d). Since layer 2 of the octupole unit cell requires negative-valued couplings in real-space, we use off-resonant link rings of a slightly smaller length $L-\eta$, similar to the construction of Hafezi et al.~\cite{hafezi_imaging_2013}. The link rings in the second layer are offset from the site rings by $\eta_- = \pi/2\beta_0$ to realize this negative coupling [Fig.~\ref{fig:4_octupole}(e)], where $\beta_0$ is the propagation constant of a mode at frequency $\omega_0$ in the waveguide forming the ring. We assume $L\gg \eta$ to ensure negligible variation of the coupling phase across the $M_\omega$ frequency dimension modes. The signs of the modulation also alternate between each neighboring site ring in the unit cell. Fig.~\ref{fig:4_octupole}(f) shows the entire 2D square array of modulated rings that realizes the octupole insulator's 3D cubic tight-binding lattice. Thus, the 2D lattice of Fig.~\ref{fig:4_octupole}(f) with a synthetic frequency dimension realizes a quantized octupole insulator supporting midgap corner-localized modes.

Using a similar recipe, we construct a hexadecapole insulator by adding a third spatial dimension to the octupole insulator, and switching the signs of all real-space couplings between vertical layers. The unit cell for the hexadecapole insulator has eight site rings with alternating signs of the modulation between adjacent site rings~[Fig.~\ref{fig:4_octupole}(g)]. Thus, we form the 4D hypercubic lattice of Fig.~\ref{fig:4_octupole}(h) in real and synthetic dimensions, supporting 0D corner modes with codimension 4, signifying a fourth-order topological insulator. Specifically, the inner cube of the hypercubic lattice in Fig.~\ref{fig:4_octupole}(h) is formed by the bottom layer of rings in Fig.~\ref{fig:4_octupole}(g), and the outer cube is realized by the top layer of rings. The two cubes are connected by positive-valued couplings in Fig.~\ref{fig:4_octupole}(h), which are implemented by the vertical coupling between the site rings in Fig.~\ref{fig:4_octupole}(g). Such vertically coupled rings have been experimentally realized routinely in silicon photonics and in III-V photonics~\cite{chu_eight-channel_1999, sherwood-droz_scalable_2011, grover_vertically_2001, yanagase_box-like_2002}. We note that the realization of the hexadecapole insulator is difficult in real space due to the three-dimensional nature of space.

\section*{Discussion}
We have introduced the concept of synthetic dimensions for realizing higher-order topological phases supporting quantized bulk quadrupole, octupole and hexadecapole moments. These phases support topologically protected zero-dimensional corner modes which are robust against disorder in the couplings. We have also shown the excitation of these corner modes in real and synthetic dimensions, and a dynamical topological phase transition between a quadrupole insulating phase and a 2D SSH phase. Future work could construct 1D boundary modes of HOTIs, such as chiral hinge states, using similar synthetic-space concepts.  Although we focused on a photonic implementation using a synthetic frequency dimension, our approach can be generalized to other degrees of freedom such as the spin or momentum of ultracold atoms and molecules, or the orbital angular momentum of light. Additional frequency dimensions can also be harnessed for this purpose~\cite{martin_topological_2017, yuan_synthetic_2018, reimer_high-dimensional_2019}. Lastly, our proposal is ripe for experimental demonstration using integrated nanophotonic platforms that can modulate resonators at frequencies approaching their FSR, especially in silicon and lithium niobate~\cite{tzuang_high_2014, zhang_broadband_2019, zhang_electronically_2019, rueda_resonant_2019}.

{\emph Note--}
While this manuscript was being prepared, we became aware of a related work using synthetic frequency and orbital angular momentum dimensions~\cite{zhang_photonic_2019}.

\section*{Acknowledgements}
This work is supported by a Vannevar Bush Faculty Fellowship (Grant No. N00014-17-1-3030) from the U. S. Department of Defense, and by MURI grants from the U. S. Air Force Office of Scientific Research (Grant No. FA9550-17-1-0002, FA9550-18-1-0379). 

\subsection*{Conflict of interests}
The authors declare no competing interests.

\section*{References}
\bibliographystyle{naturemag}
\bibliography{../../../exptdata/band_str_paper/library_2018_10_26}

\end{document}